# Arrayed van der Waals Vertical Heterostructures based on 2D GaSe Grown by Molecular Beam Epitaxy

Xiang Yuan, [1,2] Lei Tang, [1,2] Shanshan Liu,[1,2] Peng Wang,[3] Zhigang Chen,[4] Cheng Zhang,[1,2] Yanwen Liu, [1,2] Weiyi Wang,[1,2] Yichao Zou, [4] Cong Liu,[3] Nan Guo[3], Jin Zou,[4,5] Peng Zhou,[6*] Weida Hu,[3*] Faxian Xiu[1,2*]

[1]State Key Laboratory of Surface Physics and Department of Physics, Fudan University, Shanghai 200433, China

[2]Collaborative Innovation Center of Advanced Microstructures, Fudan University, Shanghai 200433, China

[3]National Laboratory for Infrared Physics, Shanghai Institute of Technical Physics, Chinese Academy of Sciences, Shanghai 200083, China

[4]Materials Engineering, The University of Queensland, Brisbane QLD 4072, Australia

[5]Centre for Microscopy and Microanalysis, The University of Queensland, Brisbane QLD 4072, Australia

[6]State Key Laboratory of ASIC and System, Department of Microelectronics, Fudan University, Shanghai 200433, China

[*]Correspondence and requests for materials should be addressed to F. X. (E-mail: faxian@fudan.edu.cn; wdhu@mail.sitp.ac.cn; pengzhou@fudan.edu.cn)

**Abstract**

Vertically stacking two dimensional (2D) materials can enable the design of novel electronic and optoelectronic devices and realize complex functionality. However, the fabrication of such artificial heterostructures in wafer scale with an atomically-sharp interface poses an unprecedented challenge. Here, we demonstrate a convenient and controllable approach for the production of wafer-scale 2D GaSe thin films by molecular beam epitaxy. *In-situ* reflection high-energy electron diffraction oscillations and Raman spectroscopy reveal a layer-by-layer van der Waals epitaxial growth mode. Highly-efficient photodetector arrays were fabricated based on few-layer GaSe on Si. These photodiodes show steady rectifying characteristics and a high external quantum efficiency of 23.6%. The resultant photoresponse is super-fast and robust with a response time of 60 μs. Importantly, the device shows no sign of degradation after 1 million cycles of operation. We also carried out numerical simulations to understand the underlying device working principles. Our study establishes a new approach to produce controllable, robust and large-area 2D heterostructures and presents a crucial step for further practical applications.

**TOC graph:**

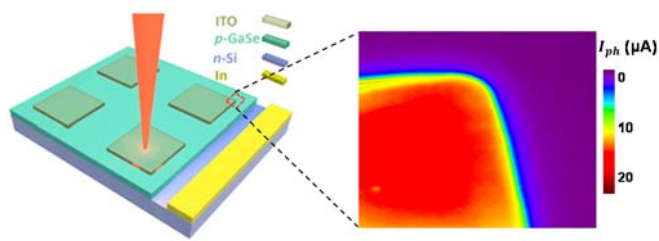

Heterostructures are ubiquitous in modern electronics and optoelectronics.[1-3] Two dimensional (2D) materials, such as graphene,[4-7] transition metal dichalcogenides (TMDs)[8-11] and GaSe,[12-15] offer opportunities for engineering novel optical and electrical devices. So far, monolayer TMDs have been demonstrated as a building block for photodetectors,[16] solar cells,[17] and avalanche photodiodes.[18] In contrast to the traditional three dimensional (3D) materials, they offer great advantages in light weight, flexibility, transparency and appealing electrical tunability.[19] In particular, significant progress has been made on $MoS_2$/Si heterostructures where a monolayer $MoS_2$ and *p*-type Si form a van der Waals heterojunction (VWH) with an external quantum efficiency (EQE) exceeding 4%.[17, 20] An avalanche photodiode[18] with the same structure shows a carrier multiplication exceeding 1000.

However, the fabrication of large-scale VWH with an atomically-sharp interface and high photoresponsivity remains challenging. Mechanical transfer technique for stacking VWH has been widely used.[16, 21] But it is not easy to precisely control stacking orientation, and the interface can also be jeopardized during the exposure to the polymers. Also, this transferring method has a low yield that is not advantageous particularly in view of the volume production of devices. In contrast, directly fabricating the heterostructures, such as through chemical vapor deposition, can produce excellent results.[22, 23] However, the small lateral size and device yield still severely hinder the real applications due to the complex fabrication process. Alternatively, molecular beam epitaxy (MBE) is well established for growing high-quality and uniform epitaxial layers. The thickness of the grown thin films can be precisely controlled by *in-situ* reflection high-energy electron diffraction (RHEED) and the interface between different layers is atomically sharp. These merits make MBE a desirable technique for fabricating wafer-scale and high-yield van der Waals heterostructures based on 2D materials. Furthermore, by integrating with the state-of-art silicon-based technology, these materials may show their potential for vast practical applications.

In this letter, we demonstrate the preparation and characterizations of vertical van der Waals heterostructures by directly depositing GaSe on *n*-type Si wafer using MBE approach. The photodiodes exhibit a fast photoresponse and prominent reversibility and stability. Even after 1 million cycles of operation, the device shows no sign of degradation. The thickness-dependent photoresponse of GaSe/Si diodes is also investigated to gain insights of the band structure evolution, which is in a good agreement with the theoretical predictions.[24]

Optimal conditions for layer-by-layer growth of GaSe thin films were established by a systematic investigation aided by *in-situ* RHEED. Figure 1A shows the RHEED intensity oscillations. Diffraction intensity was measured by a real-time image processing system. The periodic oscillations indicate a layer-by-layer 2D growth mode with a growth rate of 58 s/layer. The inset is a contrastingly streaky RHEED pattern of an 8-layer GaSe thin film, suggesting a well-ordered and atomically flat surface. All the results hereafter are obtained from the 8-layer sample unless otherwise stated. With this growth process, GaSe was uniformly deposited onto 2-inch transparent mica substrate (Figure S1). A schematic sketch of the GaSe crystal structure is depicted in Figure 1B, which is composed of several unit layers combined with van der Waals forces. Each unit layer (called tetra-layer) is consisted of four atomic layers with a stacking sequence of Se-Ga-Ga-Se (defined as one layer hereafter) and the thickness of a tetra-layer is 0.8 nm.[13, 15, 25]

The structural properties of GaSe were revealed by X-ray diffraction (XRD) and high-resolution transmission electron microscopy (HRTEM). Figure 1C is the XRD pattern, in which the diffraction peaks in blue can be indexed as rhombohedral structured GaSe phase with lattice parameters of a = b = 3.746 nm, c = 23.910 nm and a space group of R3m (PDF#65-2112).[26-28] The mica phase presented in magenta has lattice parameters of a =5.369 nm, b =9.289 nm, c =10.153 nm and a space group of C2/m (PDF#53-1188). It is of interest to note that all diffraction peaks from the GaSe correspond to the (003), (006), (0012) and (0015) planes, confirming the [0001] growth direction of GaSe thin film. For further investigation of the structure and crystal quality, GaSe were exfoliated from mica and transferred onto the TEM grid. Figure 1D shows a typical TEM image. Inset is the HRTEM image of the GaSe layer and its corresponding selected area electron diffraction pattern (SAED) taken along the [0003] zone axis, which demonstrates the high crystallinity of the GaSe layer.

One of the most distinguished features of 2D materials is their thickness-dependent property. For this reason, we explored the phonon characteristics of GaSe with different thicknesses using Raman spectroscopy. Figure 1E shows the measured Raman spectra, in which several peaks at 131.7 cm$^{-1}$ ($A_{1g}^1$), 206.1 cm$^{-1}$ ($E_{2g}^2$), 252.2 cm$^{-1}$ ($E_{1g}^2$) and 307.9 cm$^{-1}$ ($A_{1g}^2$) are consistent with the phonon vibration modes previously reported.[29] The peak located at 262.1 cm$^{-1}$ comes from the underneath mica substrate. Two A$_{1g}$ peaks are associated with the out-of-plane vibrational mode of GaSe, while E$_{1g}$ and E$_{2g}$ peaks correspond to the in-plane one.[12, 25] The peak at 252.2 cm$^{-1}$ is chosen for normalization purpose. We note that with the increase of thickness, all four peaks of GaSe become more prominent with the peak position remaining unchanged and the substrate peak intensity decreases as expected. Also, the ratio of $A_{1g}^1$ and $E_{1g}^2$ diminishes from 2.41 to 0.07 with the thickness decreasing from 30 to 3 layers. Therefore, the Raman spectroscopy can be used to identify the thickness of GaSe and be applied to other 2D materials. Energy dispersive X-ray spectrometer (EDX) was employed to detect the element composition. Figure S2 shows 1:1 atomic ratio of Ga/Se with a standard deviation less than 6% across the entire wafer.

Benefiting from the van der Waals interaction, the aforementioned growth process can be readily applied to different kinds of substrates. Here, we directly grew *p*-type GaSe thin films on 3-inch *n*-doped Si (100) substrate to form abrupt *p-n* junctions. To form a VDW heterostructures, the optimized and clean interface is crucial. Different from mica substrates which can be freshly sliced before growth, silicon needs a careful cleaning process. Here we follow the well-established RCA approach[30] to remove organic or oxide residues on the surface. The as-cleaned substrates are immediately loaded into the ultra-high vacuum chamber for proper annealing treatment before the subsequent GaSe deposition.

For GaSe prepared on silicon, the same characterizations including TEM and Raman spectroscopy were carried out and the results are displayed in Figure 2. Periodic

RHEED oscillations are still observable (Figure 2A) and show the same layer-by-layer growth mode but with a lower growth rate of ~ 2.8 min/layer. The streaky RHEED pattern also suggests well-ordered and atomically flat surface (Figure 2A inset). For further investigation of the structural properties, a thicker GaSe thin film was prepared. Figure 2B is a cross-section HRTEM image of 20-layer GaSe grown on Si, where a clear interfacial phase between GaSe and Si is observed. GaSe grows along [0001] direction as witnessed from the SAED pattern (Figure 2B inset). The thickness of a tetra-layer is determined to be 0.797 nm, identical to others.[13, 15, 25] The Raman spectra of 5-layer GaSe are completely indistinguishable on mica and Si (Figure 2C). Thus, both HRTEM and Raman reveal that the GaSe films synthesized either on insulting mica or semiconducting Si share the same crystal structure and good crystallinity, suggesting the substrate-independent epitaxial growth owing to the van der Waals interaction between layers.

We further explored the electrical properties of the GaSe/Si $p$-$n$ junctions. Indium (In) was soldered on GaSe and Si as anode and cathode, respectively (Figure 3A). The indium forms a perfect Ohmic contact with GaSe (Figure S3B). We note that there exists a negligible barrier with silicon (Figure S3C), however, this barrier does not have a major influence on the device performance. Figure 3B shows the current versus bias voltage ($I$-$V$) characteristics of a typical GaSe/Si diode at 300 K. It unveils a clear rectifying behavior with the rectification ratio over 100 under 1 V bias. By varying the bias voltage and temperature, the current rectification behavior can also be attained at low temperatures, as shown in Figure 3C. The horizontal axis, vertical axis and color gauge respectively denote the temperature, the applied bias and the absolute value of current. The current decreases with the drop of temperature which can be explained by the equation $I = I_s(\exp(\frac{qV}{k_0 T}) - 1)$, where $I_s$ is the reverse saturation current, $q$ is the electron charge, $k_0$ is the Boltzmann's constant, $V$ and $T$ are the biased voltage and temperature, respectively. The rectification ratio rapidly increases from 102 to 10934 when temperature decreases from 300 to 120 K under ±1 V. Therefore, the device can also work at low temperatures with even better performance. To verify the reproducibility of the rectifying characteristics, we randomly selected diodes across the wafer and calculated the rectification ratio through *I-V* curves. As shown in Figure 3D, the distribution of rectification ratio is centralized between 50 and 200 with 73% of the measured devices in that range.

The optoelectronic properties of these 2D *p-n* heterojunctions were further explored after confirming the repeatability of the current-rectifying characteristics. In order to achieve optimized absorption of light, 50 nm-thick transparent indium-tin oxide (ITO) array was deposited on top of GaSe as the anode electrodes (Figure 4A). Ohmic contacts are also confirmed as the prerequisite for high efficiency devices and can ensure the photocurrent exclusively coming from the junction area (Figure S5). A laser with a wavelength of 532 nm was used to illuminate the device. The spot diameter of the focused laser is 2 μm, which is much smaller than the lateral size of the junction. Figure 4B shows a spatial-resolved, zero-bias photocurrent mapping of a corner of junction under an illumination of 450 μW. We note that the photocurrent is uniform over the junction area and it decreases to zero rapidly when the laser spot moves away from the junction. Photocurrent-mapping shows sharp edges that agree well with the microscopic picture of the device. The development of the photocurrent in our patterned device arrays can be understood as follows. If there are photo-carriers generated, these carriers will recombine rapidly owing to their short life time and short diffusion length in the heavily-doped GaSe/Si materials.[31] If the photo-carriers are generated within the ITO area, most of the carriers can be collected by the ITO electrode because the photo-carriers only need to travel nanometer scale, *i.e.*, vertical transport. However, if the carriers are generated far away from the electrode, they will be most likely subjected to the recombination process before they can even reach the electrode that is hundreds of micron-meters away. Figure 4C displays the histogram of zero-biased photocurrent with a laser power of $450\,\mu W$, in which 78.9 % of diodes reach 23 μA, pointing to a good repeatability of photoresponse among the randomly-chosen diodes across the wafer.

To examine the photoresponse of the GaSe/Si heterojunctions, one of the diodes is selected and the *I-V* characteristics under different illumination conditions were acquired (Figure 4D). When the laser illuminates on the device, the negative biased current increases with increasing the incident laser power. Photocurrent is defined by the equation $I_{ph} = I_{light} - I_{dark}$. With the reverse voltage increasing, the current rises firstly and subsequently reaches saturation. This can be explained by the adequate electric field that can separate electron-hole pairs sufficiently and promptly that giving rise to the increase of the photocurrent. However, when the bias is sufficiently large, owing to the available number of charge carriers excited by the fixed illumination power, the photocurrent saturates. The dependence of $I_{ph}$ on incident power is shown in Figure 4E.

$I_{ph}$ increases with the laser illumination following a power law as a consequence of the enhanced number of excited electron-hole pairs under stronger illumination.[32, 33] EQE is defined as the number of photo-excited charge carriers per incident photon, that is, $EQE = \frac{I_{ph}}{q\phi} = \frac{I_{ph}}{q}\frac{h\upsilon}{P_{in}}$, where $I_{ph}$ is the photocurrent, $q$ is the electron charge, $\phi$ is the number of incident photon, $h$ is the Planck constant and $\upsilon$ is frequency of incident laser. Figure 4F shows the calculated EQE under various incident power at zero bias. It increases with the reduction of the incident illumination and reaches the maximum value of 23.6%, a relatively high value for 2D material without external electrical field. This value is normally smaller than 100% for photodiodes.

According to the first principle calculations, a direct-to-indirect transition for few-layer GaSe with decreasing thickness was predicted[24], distinct from the widely studied MoS$_2$ which adopts the opposite band structure transition.[34] For multiple-layer GaSe (>7 L), both the valence band maximum (VBM) and the conduction band minimum (CBM) are located at the Γ point. With the layer number decreasing (7 L), VBM splits along k space and indirect bandgap appears. As the thickness is further reduced (<7 L), the distance between VBM and CMB on the k-space increases, thus making the bandgap more "indirect". Therefore, the photo-carrier generation efficiency is expected to decrease with the film thickness. Experimentally, we performed the thickness-dependent photoresponse measurements on 2,3,5,8 layer samples at zero bias with the illumination of 532 nm laser. As shown in Figure 5, both the photocurrent and the EQE exhibit a clear power law dependence as a function of the input power, regardless of the thickness. And they decrease significantly with the layer number, presumably caused by the weakened photo-absorption in thinner samples and the band structure transition as predicted by the theory[24].

Response speed is an important criterion of photodetectors. Fast response is required for most photodetection applications. To date, the reported response time is in the range of milliseconds and minutes.[35] To study the photoresponse of our devices, we used a chopper to periodically switch the laser on and off and the photocurrent was then recorded on an oscilloscope with a preamplifier. Figure 6A shows one complete circle of photoresponse at zero bias and under 450 μW of 532 nm laser. As shown in Figure 6A, when laser turns on, the current increases quickly and reaches saturation at 23.1 μA. Once the laser turns off, the current decreases rapidly to zero. Photoresponse time, defined as the switching duration from 10% to 90% of the maximum photocurrent, was calculated as 60 μs and 20 μs for the increasing/falling edges, respectively. The fast response speed demonstrated in our GaSe/Si photodiodes promises further practical applications. In Figure 6B, after 1,000,000 cycles of operation, *i.e.*, laser switching on and off, the photoresponse behavior retains almost the same as the initial states, suggesting an outstanding reversibility and stability of GaSe/Si photodetectors.

In order to further understand the device performance, numerical simulations were carried out. The carrier concentration of GaSe and Si is extracted respectively to be $3.1\times10^{18}$ cm$^{-3}$ and $6.4\times10^{18}$ cm$^{-3}$ from the Hall effect measurements (Figure S4). Figure 7 shows the energy-band diagrams of the *p-n* junction under a voltage bias ($V=\pm 1$ V) and the thermal equilibrium condition (0 V). At equilibrium, an one-sided abrupt junction or $n^+p$ heterojunction is formed as the carrier concentration in the heterojunction changes abruptly from electrons of $N_e = 6.4\times10^{18}$ cm$^{-3}$ to holes of $N_h = 3.1\times10^{18}$ cm$^{-3}$, as shown in the middle panel of Figure 7.

According to the simulation results, under the thermal equilibrium (0 V), more than 60% of the GaSe epitaxial layer (5 among 8 layers) is depleted, which makes the few-layer GaSe a good photo-absorption and carrier-generating material. While negative 1V bias is employed, the depletion region extends to nearly 90% (7 among 8 layers), resulting in the increase of the photocurrent as shown in Fig. 4d. On the other hand, the depletion region is almost reduced to zero (0 among 8 layers) at the positive 1V bias which is in good agreement with the high rectifying ratio (Fig. 3b and 3d). More than half of the thickness is depleted, giving rise to the high quantum efficiency at the thermal equilibrium and better performance at the negative bias. The fast speed of the device can also be interpreted. The highly depleted region in GaSe leads to a strong built-in electric field which efficiently separates the photo-generating carriers causing the high speed.[31] Furthermore, the truly vertical contacts reduce the transport distance to the nanometer scale. Not only it increases the efficiency due to the reduced undesirable recombination, but it also dramatically enhances the response speed.

In summary, we showed a controllable growth method for producing wafer-scale 2D materials by MBE. Arrayed van der Waals vertical heterostructures were achieved based on few-layer GaSe on Si. The heterostructures can work as *p-n* diodes with steady rectifying behavior and highly efficient, fast, robust photoresponse. The wafer-scale size and striking performance of the heterostructures make GaSe/Si promising for real applications. Integrating atomically-thin 2D materials with traditional materials is a stirring choice for next-generation optoelectronics.

## Methods

**Thin film synthesis.**

Layered GaSe was grown on substrates of freshly sliced mica or RCA (Radio Corporation of America) processed Si in Perkin Elmer 430 MBE system. High-purity

Ga and Se were evaporated from standard Knudsen cells. Optimal growth conditions were examined by *in-situ* RHEED system and the growth was conducted under the flux ratio of Se/Ga 10:1 which was measured by a crystal oscillator. The growth temperature was 580 ℃.

**Structural characterizations.**

Raman spectra were measured with an incident laser of 633 nm (Renishaw-inVia confocal Raman system). Crystal quality of GaSe thin films was investigated by XRD (Bruker D8 Discover) and HRTEM (FEI Tecnai F20).

**Electrical and optoelectrical characterizations.**

Electrical characteristics and photoresponse measurements were carried out using Agilent 2902. The incident light from a 532 nm-laser was focused on the samples. The intensity was controlled by a series of neutral density filters plus a linear polarizer/wave-plate and calibrated by Thorlab-S130C photodiode power sensor. For the time-resolved photocurrent measurements, the laser was controlled by Thorlab-ITC4001 controller, working at chopper frequency of 761 Hz. The Tektronix MDO 3014 oscilloscope was used to read the signals. For the spatial-resolved photocurrent measurements, a galvanometer were controlled by an open-source controller, thus allowing the laser spot scanning across the device.

**Numerical Simulations.**

To investigate the energy-band diagrams in the $n^+p$ heterojunction under different bias conditions, 2D steady-state numerical simulations were performed using the Sentaurus Device, a commercial software package from Synopsys. For the drift-diffusion simulations, the well-known Poisson and continuity equations are used in the calculations. The carrier generation-recombination process consists of Shockley-Read-Hall, Auger, and radiative terms. Additionally, tunneling effects, such as band-to-band and trap-assisted tunneling, are included in the simulations.

**Associated Content**

Supporting Information.

Description of experimental results: Sample photo, EDX spectrum, contact performance, Hall Effect measurements. This material is available free of charge via the Internet at http://pubs.acs.org.

**Acknowledgment**

This work was supported by the National Young 1000 Talent Plan, Pujiang Talent Plan in Shanghai, and National Natural Science Foundation of China (61322407, 11474058, 11322441). Part of the sample fabrication was performed at Fudan Nano-fabrication Laboratory. We thank Yizheng Wu, Boliang Chen, Xiao Yan for great assistance during the device fabrication and measurements.

# REFERENCES

1. Kroemer, H. *Proceedings of the IEEE* **1982,** 70, (1), 13-25.
2. Ohno, Y.; Young, D.; Beschoten, B. a.; Matsukura, F.; Ohno, H.; Awschalom, D. *Nature* **1999,** 402, (6763), 790-792.
3. Geim, A. K.; Grigorieva, I. V. *Nature* **2013,** 499, (7459), 419-25.
4. Bonaccorso, F.; Sun, Z.; Hasan, T.; Ferrari, A. C. *Nature Photonics* **2010,** 4, (9), 611-622.
5. Mueller, T.; Xia, F.; Avouris, P. *Nature Photonics* **2010,** 4, (5), 297-301.
6. Xia, F.; Mueller, T.; Lin, Y. M.; Valdes-Garcia, A.; Avouris, P. *Nature nanotechnology* **2009,** 4, (12), 839-43.
7. Tian, H.; Yang, Y.; Xie, D.; Cui, Y. L.; Mi, W. T.; Zhang, Y.; Ren, T. L. *Scientific reports* **2014,** 4, 3598.
8. Wang, Q. H.; Kalantar-Zadeh, K.; Kis, A.; Coleman, J. N.; Strano, M. S. *Nat Nanotechnol* **2012,** 7, (11), 699-712.
9. Radisavljevic, B.; Kis, A. *Nat Mater* **2013,** 12, (9), 815-20.
10. Lopez-Sanchez, O.; Lembke, D.; Kayci, M.; Radenovic, A.; Kis, A. *Nat Nanotechnol* **2013,** 8, (7), 497-501.
11. Ross, J. S.; Klement, P.; Jones, A. M.; Ghimire, N. J.; Yan, J.; Mandrus, D. G.; Taniguchi, T.; Watanabe, K.; Kitamura, K.; Yao, W.; Cobden, D. H.; Xu, X. *Nat Nanotechnol* **2014,** 9, (4), 268-72.
12. Hu, P.; Wen, Z.; Wang, L.; Tan, P.; Xiao, K. *ACS nano* **2012,** 6, (7), 5988-5994.
13. Lei, S.; Ge, L.; Liu, Z.; Najmaei, S.; Shi, G.; You, G.; Lou, J.; Vajtai, R.; Ajayan, P. M. *Nano Lett* **2013,** 13, (6), 2777-81.
14. Mahjouri-Samani, M.; Gresback, R.; Tian, M.; Wang, K.; Puretzky, A. A.; Rouleau, C. M.; Eres, G.; Ivanov, I. N.; Xiao, K.; McGuire, M. A.; Duscher, G.; Geohegan, D. B. *Advanced Functional Materials* **2014,** 24, (40), 6365-6371.
15. Zhou, Y.; Nie, Y.; Liu, Y.; Yan, K.; Hong, J.; Jin, C.; Zhou, Y.; Yin, J.; Liu, Z.; Peng, H. *ACS nano* **2014,** 8, (2), 1485-1490.
16. Hong, X.; Kim, J.; Shi, S. F.; Zhang, Y.; Jin, C.; Sun, Y.; Tongay, S.; Wu, J.; Zhang, Y.; Wang, F. *Nat Nanotechnol* **2014,** 9, (9), 682-6.
17. Tsai, M.-L.; Su, S.-H.; Chang, J.-K.; Tsai, D.-S.; Chen, C.-H.; Wu, C.-I.; Li, L.-J.; Chen, L.-J.; He, J.-H. *ACS nano* **2014,** 8, (8), 8317-8322.
18. Lopez-Sanchez, O.; Dumcenco, D.; Charbon, E.; Kis, A. *arXiv preprint arXiv:1411.3232* **2014**.
19. Xia, F.; Wang, H.; Xiao, D.; Dubey, M.; Ramasubramaniam, A. *Nature Photonics* **2014,** 8, (12), 899-907.
20. Lopez-Sanchez, O.; Alarcon Llado, E.; Koman, V.; Fontcuberta i Morral, A.; Radenovic, A.; Kis, A. *ACS nano* **2014,** 8, (3), 3042-3048.
21. Deng, Y.; Luo, Z.; Conrad, N. J.; Liu, H.; Gong, Y.; Najmaei, S.; Ajayan, P. M.; Lou, J.; Xu, X.; Ye, P. D. *ACS nano* **2014,** 8, (8), 8292-8299.
22. Gong, Y.; Lin, J.; Wang, X.; Shi, G.; Lei, S.; Lin, Z.; Zou, X.; Ye, G.; Vajtai, R.; Yakobson, B. I. *Nature materials* **2014,** 13, (12), 1135-1142.
23. Duan, X.; Wang, C.; Shaw, J. C.; Cheng, R.; Chen, Y.; Li, H.; Wu, X.; Tang, Y.; Zhang, Q.; Pan, A.; Jiang, J.; Yu, R.; Huang, Y.; Duan, X. *Nat Nano* **2014,** 9, (12), 1024-1030.
24. Li, X.; Lin, M.-W.; Puretzky, A. A.; Idrobo, J. C.; Ma, C.; Chi, M.; Yoon, M.; Rouleau, C. M.; Kravchenko, I. I.; Geohegan, D. B. *Scientific reports* **2014,** 4.
25. Mahjouri-Samani, M.; Tian, M.; Wang, K.; Boulesbaa, A.; Rouleau, C. M.; Puretzky, A. A.; McGuire,

M. A.; Srijanto, B. R.; Xiao, K.; Eres, G. *ACS nano* **2014,** 8, (11), 11567-11575.

26. Madelung, O., IIIx-VIy compounds. In *Semiconductors: Data Handbook*, Springer: 2004; pp 515-552.

27. KARABULUT, M.; ERTAP, H.; MAMMADOV, H.; UĞURLU, G.; ÖZTÜRK, M. K. *Turkish Journal of Physics* **2014,** 38, (1).

28. Jurca, H. F.; Mazzaro, I.; Schreiner, W. H.; Mosca, D. H.; Eddrief, M.; Etgens, V. H. *Thin Solid Films* **2006,** 515, (4), 1470-1474.

29. Balitskii, O. A.; Borowiak-Palen, E.; Konicki, W. *Crystal Research and Technology* **2011,** 46, (4), 417-420.

30. Kern, W. *New Jersey: Noyes Publication* **1993**, 111-196.

31. Sze, S. M., *Semiconductor devices: physics and technology*. John Wiley & Sons: 2008.

32. Koster, L.; Mihailetchi, V.; Ramaker, R.; Blom, P. *Applied Physics Letters* **2005,** 86, (12), 123509-123509-3.

33. Tsai, H.-L.; Tu, C.-S.; Su, Y.-J. In *Development of generalized photovoltaic model using MATLAB/SIMULINK*, Proceedings of the world congress on Engineering and computer science, 2008; pp 1-6.

34. Splendiani, A.; Sun, L.; Zhang, Y.; Li, T.; Kim, J.; Chim, C.-Y.; Galli, G.; Wang, F. *Nano letters* **2010,** 10, (4), 1271-1275.

35. Sun, Z.; Chang, H. *ACS nano* **2014**.

**Figures**

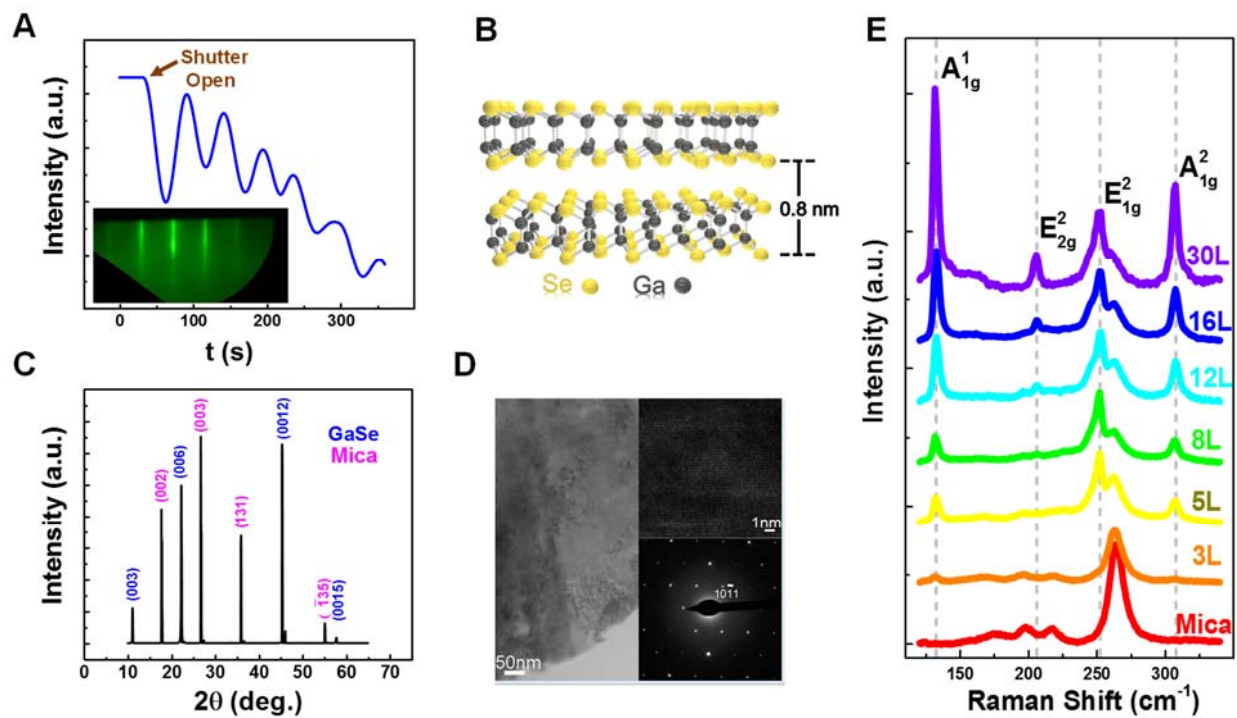

Figure 1. Synthesis and characterizations of 8-layer GaSe deposited on mica. (A) RHEED oscillations. Periodic oscillations can function as adequate evidence of layer-by-layer growth. The growth rate is 58 s/layer. The inset is a streaky RHEED pattern of GaSe film indicating well-ordered and atomically flat surface. (B) Layered crystal structure of GaSe with each tetra-layer formed by Se-Ga-Ga-Se. (C) A distinctive XRD pattern of GaSe. Marked peaks (in blue) show GaSe's characteristic peaks while others (in magenta) result from mica substrate. (D) TEM, HRTEM and SAED pattern of the exfoliated GaSe flake. (E) Thickness-dependent Raman spectra. The intensity ratio of $A_{1g}^1$ and $E_{1g}^2$ can be used as a probe to determine the film thickness.

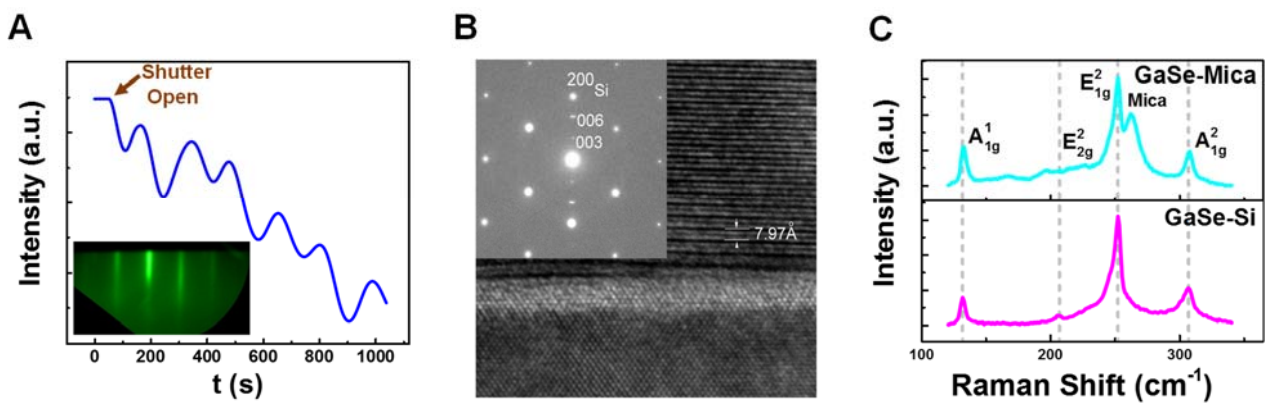

Figure 2. Synthesis and characterizations of GaSe thin films deposited on Si. (A) REHEED oscillations. It shows periodic oscillations and the growth rate is 2.8 min/layer. The inset is a streaky RHEED pattern of 8-layer epitaxial GaSe film. (B) HRTEM cross-section image of 20-layer GaSe/Si and the SAED pattern (inset). (C) Raman spectra of GaSe grown on mica (upper panel) and Si (lower panel). The characteristic peaks are exactly the same indicating substrate-independent growth with similar crystal quality owing to the van der Waals nature of 2D materials.

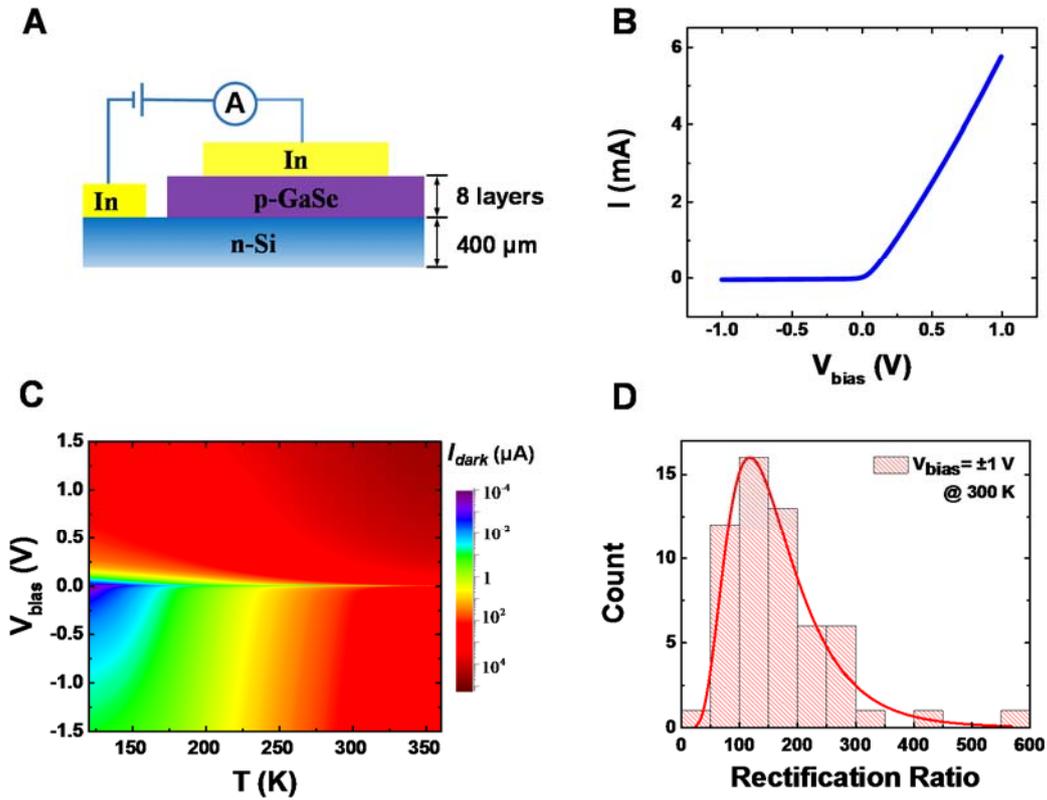

Figure 3. Electrical characteristics of 8-layer GaSe/Si *p-n* junctions. (A) A schematic sketch of the device. (B) Current vs bias voltage characteristic of the GaSe/Si heterojunction diode. Clear rectification behavior with rectification ratio higher than 100 is achieved. (C) 2D plot of dark current ($I_{dark}$) as a function of temperature and bias voltage without light illumination. The horizontal axis, vertical axis and color gauge denote the temperature, the applied bias and the current, respectively. The rectification ratio rapidly increases from 102 to 10934 when temperature decreases from 300 to 120 K. (D) A bar graph of rectification ratio distribution from randomly-selected diodes across 3-inch GaSe/Si wafer, indicative of a good repeatability.

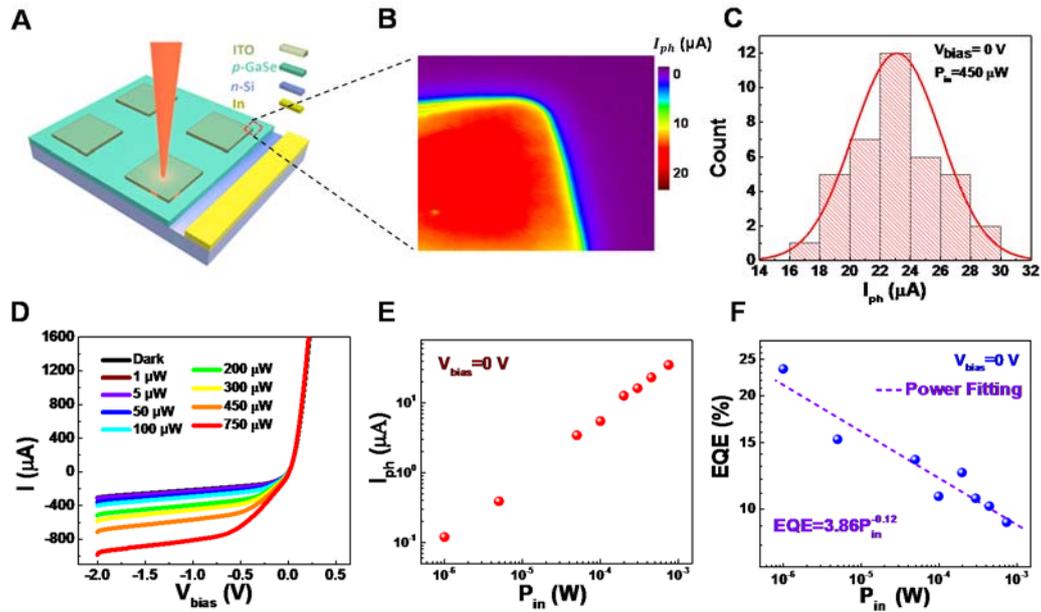

Figure 4. Photoresponse of 8-layer GaSe/Si *p-n* junctions. (A) A schematic diagram of the device structure. (B) A mapping of the spatial-resolved, zero-bias photocurrent with laser situated at a corner of junction area under the illumination of 450 μW. The photocurrent distributes uniformly in the junction area. Also, the well-defined mapping edge can verify the vertical structure of the device. (C) A bar graph of photocurrent distribution measured at zero bias based on randomly-chosen diodes on 3-inch GaSe/Si wafer. (D) The negative biased *I-V* characteristics of the *p-n* diode under various incident laser power. The inset shows a clear *I-V* curve around zero bias. (E) zero-biased photocurrent under different incident intensity. (F) EQE under different incident power at zero bias. With the decrease of laser intensity, the EQE rises up to 23.6% at 1 μW.

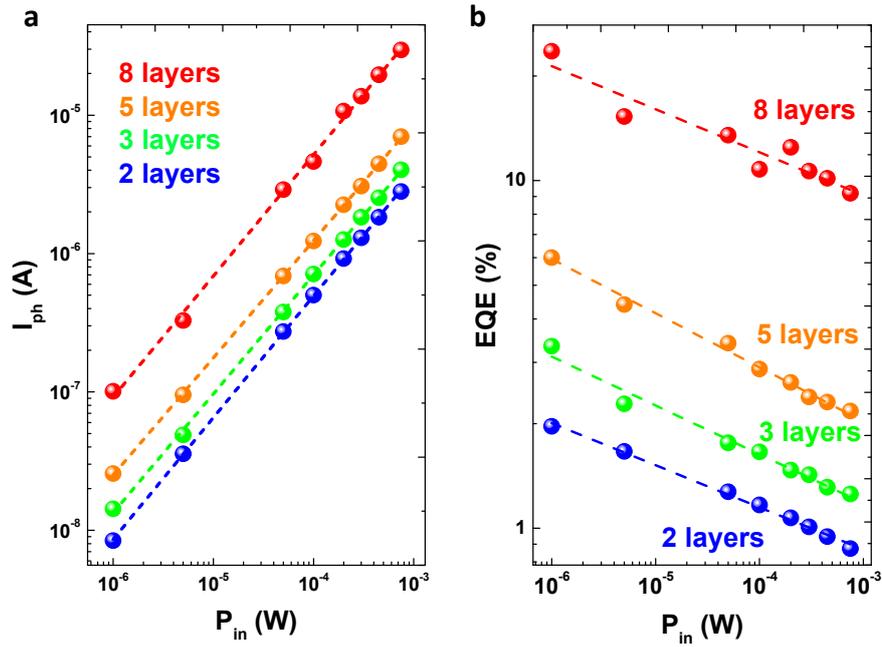

Fig. 5. Thickness-dependent photoresponse. (A) The zero-bias thickness-dependent photocurrent under 532.8 nm laser illumination with different incident power. The dashed line is the power-law fitting. (B) The external quantum efficiency under different incident power. The dashed line is the power-law fitting.

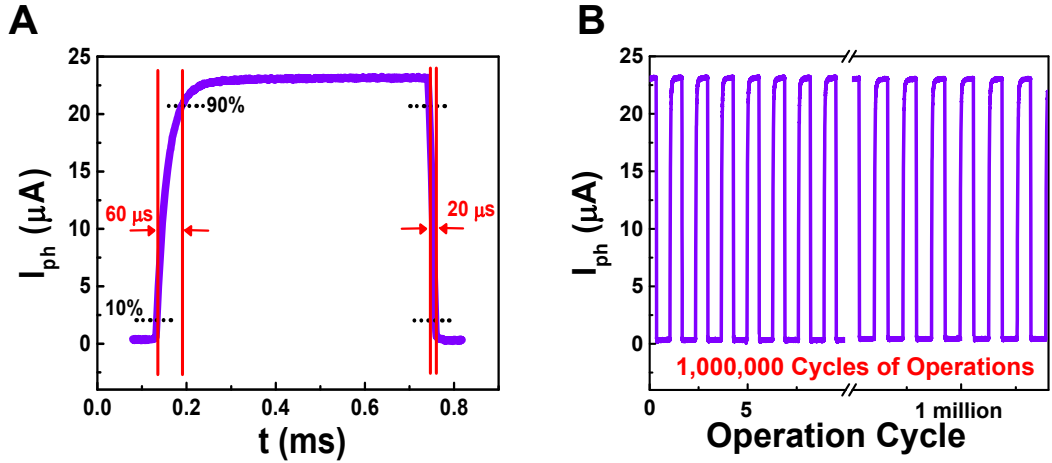

Figure 6. Time-resolved photoresponse and stability test of the 8-layer GaSe/Si device. (A) A cycle of photo-switching photocurrent at zero bias under 450 μW. When the laser turns on, photocurrent increases sharply and the rising photoresponse time is 60 μs. When the laser is off, the photocurrent decreases to zero promptly and the decay time is 20 μs. (B) Stability test of device. Under 1 million cycles of on/off operation, the photoresponse retains unchanged.

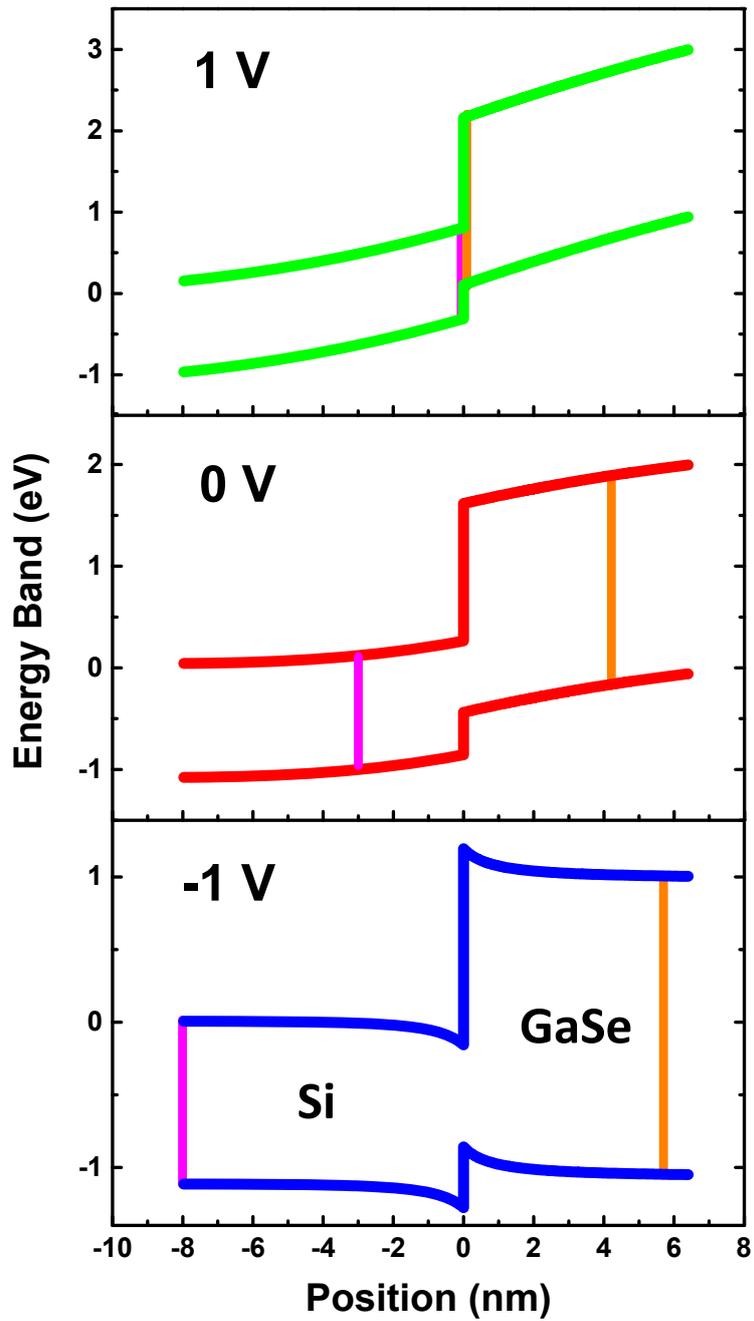

Figure 7. Numerical Simulations of the band diagrams under different bias. The spacing between two vertical lines in both materials denotes the depletion depth.